\documentclass[a4paper]{llncs}

\makeatletter
\def\hlinewd#1{%
  \noalign{\ifnum0=`}\fi\hrule \@height #1 \futurelet
   \reserved@a\@xhline}
\makeatother

\usepackage{mathrsfs}
\usepackage{amssymb}
\usepackage{graphicx}
\usepackage{subfigure}
\usepackage{multirow}
\usepackage{threeparttable}
\usepackage{CJK}

\usepackage{url}
\urldef{\mailsa}\path|{jiaming.xu, boxu, guanhua.tian, fangyuan.wang, hongwei.hao}@ia.ac.cn,{jzhao}@nlpr.ia.ac.cn|
\newcommand{\keywords}[1]{\par\addvspace\baselineskip
\noindent\keywordname\enspace\ignorespaces#1}

\usepackage{algorithm}
\usepackage{algorithmic}

\usepackage{amssymb,  subfigure}

\begin{document}

\mainmatter  
\title{Short Text Hashing Improved by Integrating Multi-Granularity Topics and Tags}

%
%
%


%
%
\author{Jiaming Xu%
\and Bo Xu\and Guanhua Tian\and Jun Zhao\and\\
Fangyuan Wang\and Hongwei Hao}
\authorrunning{J.M. Xu et al.}
\institute{Institute of Automation, Chinese Academy of Sciences. 100190, Beijing, P.R. China\\
\mailsa\\
}

%
%

\maketitle

\begin{abstract}
Due to computational and storage efficiencies of compact binary codes, hashing has been widely used for large-scale similarity search. Unfortunately, many existing hashing methods based on observed keyword features are not effective for short texts due to the sparseness and shortness. Recently, some researchers try to utilize latent topics of certain granularity to preserve semantic similarity in hash codes beyond keyword matching. However, topics of certain granularity are not adequate to represent the intrinsic semantic information. In this paper, we present a novel unified approach for {\em short text Hashing using Multi-granularity Topics and Tags}, dubbed HMTT. In particular, we propose a selection method to choose the optimal multi-granularity topics depending on the type of dataset, and design two distinct hashing strategies to incorporate multi-granularity topics. We also propose a simple and effective method to exploit tags to enhance the similarity of related texts. We carry out extensive experiments on one short text dataset as well as on one normal text dataset. The results demonstrate that our approach is effective and significantly outperforms baselines on several evaluation metrics.

\end{abstract}

\section{Introduction}
\label{sec:Introduction}
With the explosion of social media, numerous short texts become available in a variety of genres, e.g. tweets, instant messages, questions in Question and Answer (Q\&A) websites and online advertisements~\cite{Transferring_Jin_2011}. In order to conduct fast similarity search in those massive datasets, hashing, which tries to learn similarity-preserving binary codes for document representation, has been widely used to accelerate similarity search.
Unfortunately, many existing hashing methods based on keyword feature space usually fail to fully preserve the semantic similarity of short texts due to the sparseness of the original feature space. For example, there are three short texts as follows:

  \(d1\): ``{\em Rafael Nadal missed the Australian Open}'';

  \(d2\): ``{\em Roger Federer won Grand Slam title}'';

  \(d3\): ``{\em Tiger Woods broke numerous golf records}''.

Obviously, the hashing methods based on keyword space cannot see the similarity among \(d1\), \(d2\) and \(d3\). In recent years, some researchers seek to address the challenge by latent semantic approach. For example, Wang et al.~\cite{SHTTM_Wang_2013} preserve the semantic similarity of documents in hash codes by fitting the topic distributions, and Xu et al.~\cite{FastMatch_Xu_2013} directly treat the latent topic features as tokens to represent one document for hashing learning.
However, topics of certain granularity are not adequate to represent the intrinsic semantic information~\cite{ShorTextMultiTopic_Chen_2011}. As we know, different topic models with pre-defined number of topics can extract different semantic level topics. For example, the topic model with a large number of topics can extract more fine grained topic features, such as ``Tennis Open Progress'' for \(d1\) and \(d2\), and ``Golf Star News'' for \(d3\), but fail to construct the semantic relevance of \(d3\) with the other texts, and the topic model with a few topics can extract more coarse grained semantic features, such as ``Sport'' and ``Star'' for \(d1\), \(d2\) and \(d3\), but lack distinguishing information and cannot learn the hashing function effectively, As a reasonable assumption, multi-granularity topics are more suitable to preserve semantic similarity and learn hashing function for short text hashing.

On the other hand, tags are not fully utilized in many hashing methods. Actually, in various real-world applications, documents are often associated with multiple tags, which provide useful knowledge in learning effective hash codes~\cite{SHTTM_Wang_2013}. For instance, in Q\&A websites, each question has category labels or related tags assigned by its questioner. Another example is microblog, some tweets are labeled by their authors with hashtags in the form of ``\#keyword''. Thus, we should fully exploit the information contained in tags to strengthen the semantic relationship of related texts for hashing learning.

Based on the above observations, this paper proposes a unified {\em short text Hashing using Multi-granularity Topics and Tags}, referred as HMTT for simplicity. In HMTT, two different ways are introduced to incorporate multi-granularity topics and tag information for improving short text hashing.

The main contributions of this paper are three-fold:
Firstly, a novel unified short text hashing is proposed. To our best knowledge, this is the first time of incorporating multi-granularity topics and tags into a unified hashing approach, and experiments are conducted to verify our assumption that short text hashing can be improved by integrating multi-granularity topics and tags.
Secondly, the optimal multi-granularity topics can be selected automatically, i.e., to extract effective latent topic features for hashing learning. The experimental results indicate the optimal multi-granularity topics can achieve better performances, compared with other multi-granularity topics.
Finally, two strategies to incorporate multi-granularity topics for short text hashing are designed and compared through extensive experimental evaluations and analyses.


\section{Related Work}
\label{sec:RelatedWork}
Hash-based methods can be mainly divided into two categories. One category is data-oblivious hashing. As the most popular hashing technique, Locality-Sensitive Hashing (LSH)~\cite{LSH_Andoni_2006}~based on random projection has been widely used for similarity search. However, since they are not
aware of data distribution, those methods may lead to generate quite inefficient hash codes in practice~\cite{STHs_Zhang_2010}. Recently, more researchers focus attention on the other category, data-aware hashing,
For example, the Spectral Hashing (SpH)~\cite{SpH_Weiss_2008} generates compact binary codes by forcing the balanced and uncorrelated constraints into the learned codes. Self-Taught Hashing (STH)~\cite{STH_Zhang_2010} and Two Step Hashing (TSH)~\cite{TwoStepHash_Lin_2013} decompose the learning procedure into two steps: generating binary code and learning hash function, and a supervised version of STH is proposed in~\cite{STHs_Zhang_2010} denoted as STHs.
However, the previous hashing methods, directly working in keyword feature space, usually fail to fully preserve semantic similarity. More recently, Wang et al.~\cite{SHTTM_Wang_2013} proposed a Semantic Hashing using Tags and Topic Modeling (SHTTM). However, the limitations of SHTTM are that: Although the topic distributions are used to preserve the content similarity to generate hash codes, they do not utilize the topics to improve hashing function learning; Even the number of topics must keep consistent with dimensions of hash code, that this assumption is too strict to capture the optimal semantic features for different types of datasets.

\begin{figure*}[t]
\begin{center}
\includegraphics[width=12cm]{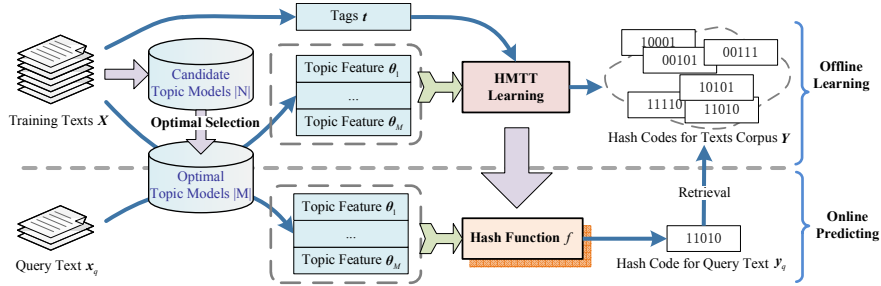}
\vspace{-0.2cm}
\caption{The proposed approach HMTT for short text hashing}\label{fig:approachoverview}
\end{center}
\vspace{-0.3cm}
\end{figure*}
\section{Algorithm Description}
\label{sec:AlgorithmDescription}
A unified short text hashing approach HMTT is depicted in Fig.~\ref{fig:approachoverview}. Given a dataset of \(n\) training texts denoted as: \({  \bf{X}} =   \{ {{\bf{x}}_1},{{\bf{x}}_2},...,{{\bf{x}}_n}\}    \in   {\mathbb{R}^{d \times n}}\), where \(d\) is the dimensionality of the
keyword feature. Denote their tags as: \(  {\bf{t}}   =   \{ {{\bf{t}}_1},{{\bf{t}}_2},...,{{\bf{t}}_n}\}    \in   {\{ 0,1\} ^{q \times n}}\), where \(q\) is the total number of possible tags associated with each text. A tag with label 1 means a text is associated with a certain tag/category, while a tag with label 0 means a missing tag or the text is not associated with that tag/category. The goal of HMTT is to obtain optimal binary codes \({ \bf{Y}  } =   {\{ {{\bf{y}}_1},{{\bf{y}}_2},...,{{\bf{y}}_n}\} ^T}   \in   {\{  - 1,1\} ^{n \times l}}\), and a hashing function \(f\): \({  \mathbb{R}^d}   \to   {\{  - 1,1\} ^l}\), which embeds the query text \({{\bf{x}}_q}\) to its binary vector representation \({{\bf{y}}_q}\) with \(l\) bits. To achieve the similarity-preserving property, we require the similar texts to have similar binary codes in Hamming space. We first select the optimal topic models from the candidate topic models, and extract the multi-granularity topic features \(\{ {{\vec{\theta }}_1},{{\vec{\theta }}_2},...,{{\vec{\theta }}_M}\} \). Then the binary codes and hash functions can be learned by integrating multi-granularity topic features and tags. In the second phase which is online, the query text is represented by binary code mapped from the derived hash function, and then the approximate nearest neighbor search is accomplished in Hamming space. All pairs of hash code found within a certain Hamming distance of each other are semantic similar texts.

The main challenges of the idea are that: (1). How to select the optimal topic models; (2). How to utilize the tag information efficiently; and (3). How to integrate the multi-granularity topics to preserve semantic similarity. The proposed approach HMTT will be described in detail in the following sections.

\begin{algorithm}[t]\caption{The Optimal Topics Selection}\label{alg:weightTopic}
\begin{algorithmic}[1]
\renewcommand{\algorithmicrequire}{\textbf{Input:}}
\renewcommand{\algorithmicensure}{\textbf{Output:}}
\REQUIRE \
\(n\) training texts \({\bf{X}}   =   \{ {{\bf{x}}_1},{{\bf{x}}_2},...,{{\bf{x}}_n}\} \) with tags \({\bf{t}}   =   \{ {{\bf{t}}_1},{{\bf{t}}_2},...,{{\bf{t}}_n}\} \), \(N\) candidate topic sets \({\bf{T}}   =   \{ {T_1},{T_2},...,{T_N}\} \) and a specified number \(M\).
\ENSURE \
The optimal topic sets \({\bf{O}}\), and the weight vector \({\vec{ \mu }}\).
\STATE Sample a sub-set \({\bf{\hat X}}\) with tags \({\bf{\hat t}} \); Initialize \({\vec{\mu }} \leftarrow {\bf{0}}\), and \({\bf{O}}   \leftarrow  \emptyset \);
\FOR {each text \({\bf{\hat x}}   \in   {\bf{\hat X}}\)}
\STATE Find \(n{n^ + }({\bf{\hat x}})\) and \(n{n^ - }({\bf{\hat x}})\);
\FOR {\(i   \leftarrow   1{\rm{ ~to~ }}N\)}
\STATE Update \(\mu ({T_i})\) by Eq.~\ref{eq:updateWeightU};
\ENDFOR
\ENDFOR
\WHILE {\(size({\bf{O}})   <   M\)}
\STATE \({T^{(p)}}   =   \arg {\max _{{T_i} \in {\bf{T}}}}{\bf{\mu }}({T_i})\); Update \({\bf{O}}   =   {\bf{O}}   \cup   \{ {T^{(p)}}\} \), \({\bf{T}}   =   {\bf{T}}   -   \{ {T^{(p)}}\} \);
\ENDWHILE
\RETURN \({\bf{O}}\) and \({\vec{ \mu }}\);
\end{algorithmic}
\end{algorithm}
\vspace{-0.3cm}

\subsection{Estimate and Select the Optimal Topics}
\label{subsec:EstimateandSelectTopic}
\label{subsubsec:EstimateCandidateTopic}
In this work, we straightforwardly obtain a set of candidate topics by pre-defining several different topic numbers of Latent Dirichlet Allocation (LDA)~\cite{LDA_Blei_2003}. After training the topic models, we can draw multi-granularity topic features, corresponding as distributions over the topics, from the candidate topic models.
\label{subsubsec:SelectOptimalTopic}

In order to select the optimal topic models, we should utilize the tag information to evaluate the quality of topics. Inspired by~\cite{ShorTextMultiTopic_Chen_2011,Estimate_Kononenko_1994}, the selection of optimal topic model sets depends on their capability in helping discriminate short texts without sharing any common tags.
We denote \(N\) different sets of topics as \({\bf{T}}   =   \{ {T_1},{T_2},...,{T_N}\} \). For each entry \({T_i}\), the probability topics distributions over documents are denoted as \({\vec {\theta}}    =   p({\bf{z}}|{\bf{x}})\). The weight vector is \({\vec{\mu }}   =   \{ \mu ({T_1}),\mu ({T_2}),...,\mu ({T_N})\} \), where \(\mu ({T_i})\) is the weight indicating the importance of topic set. The purpose is to select the optimal topic sets \({\bf{O}}   =   \{ {T_1},{T_2},...,{T_M}\} \). In~\cite{ShorTextMultiTopic_Chen_2011}, Chen et al. evaluate the quality of topics based on two aspects: discrimination and complementarity of the multi-granularity topics. However, how to balance those two aspects is a tricky problem and the latter aspect, complementarity, is easy to introduce noises for preserving similarity. Thus, we propose a simple and effective method directly based on the key idea of Relief~\cite{Estimate_Kononenko_1994} as follows: Firstly, a sub-set \({\bf{\hat X}}   =   \{ {{\bf{\hat x}}_1},{{\bf{\hat x}}_2},...,{{\bf{\hat x}}_m}\} \) with tags \({\bf{\hat t}}   =   \{ {{\bf{\hat t}}_1},{{\bf{\hat t}}_2},...,{{\bf{\hat t}}_m}\} \) is sampled from training dataset, and we find two groups of \(k\) nearest neighbors for each text \({{\bf{\hat x}}_i}\): one group is from the texts sharing any common tags (denoted as \(n{n^ + }({\bf{\hat x}})\)), and the other from the texts not sharing any common tags (denoted as \(n{n^ - }({\bf{\hat x}})\)). Then the weight is updated as follows:
\begin{equation}
\begin{array}{l}
\mu ({T_i}) = \mu ({T_i}) + \sum\limits_{j = 1}^{{k }} {\frac{{{D_{ K L}}({T_i}({\bf{x}}),{T_i}(nn_j^ - ({\bf{x}})))}}{{{k }}}} - \sum\limits_{p = 1}^{{k }} {\frac{{{D_{ K L}}({T_i}({\bf{x}}),{T_i}(nn_p^ + ({\bf{x}})))}}{{{k }}}}
\end{array}
\label{eq:updateWeightU}
\end{equation}
where, \({D_{ K L}}\) is the symmetric Kullback-Leibler (KL) divergence:
\begin{displaymath}
\begin{array}{l}
{D_{ K L}} ({T_i}({\bf{x}}), {T_i}(nn_j^ -  ({\bf{x}})))   =   \frac{1}{2}    \sum\limits_{{z_k} \in {T_i}}     {(p({z_k}|{\bf{x}})   \cdot   log} ( \frac{{p({z_k}|{\bf{x}})}}{{p({z_k}|nn_j^ - ({\bf{x}}))}} ) \\
\quad\quad\quad\quad\quad\quad\quad\quad\quad\quad\quad\quad + p({z_k}|nn_j^ - ({\bf{x}}))   \cdot   log (\frac{{p({z_k}|nn_j^ - ({\bf{x}}))}}{{p({z_k}|{\bf{x}})}})),
\end{array}
\end{displaymath}
so is the value of \({D_{ K L}}({T_i}({\bf{x}}),{T_i}(nn_p^ + ({\bf{x}})))\).
After updating the weight vector, we directly select the optimal topic sets \({\bf{O}}\) according to the top-\(M\) weight values. In summary, the optimal topics selection procedure is depicted in Algorithm~\ref{alg:weightTopic}.

\subsection{Content Similarity and Tags Preservation}
\label{subsec:TagPreserve}
In hashing problem, one key component is how to define the affinity matrix \({\bf{S}}\). Diverse approaches can be applied to construct the similarity matrix. In this paper, we choose cosine function as an example and use the local similarity structure of all text pairs to reconstruct the similarity function as follows:
\begin{equation}
\label{eq:localSimialrity}
{S_{ij}}  =  \left\{   {\begin{array}{*{20}{c}}
{{c_{ij}} \cdot \frac{{{\bf{x}}_i^T{{\bf{x}}_j}}}{{\left\| {{{\bf{x}}_i}} \right\| \cdot \left\| {{{\bf{x}}_j}} \right\|}},}&{    if \, {{\bf{x}}_i} \in {\bf{N}}{{\bf{N}}_k}({{\bf{x}}_j}) \, or \, vice \, versa}\\
0,&{otherwise}
\end{array}} \right.
\end{equation}
where \({\bf{N}}{{\bf{N}}_k}({\bf{x}})\) represents the set of \(k\)-nearest-neighbors of \({\bf{x}}\), and \({c_{ij}}\) is an confidence coefficient. If two documents \({{\bf{x}}_i}\) and \({{\bf{x}}_j}\) share any common tag, we set \({c_{ij}}\) a higher value \(a\). In reverse, the \({c_{ij}}\) is given a lower value \(b\) if two documents \({{\bf{x}}_i}\) and \({{\bf{x}}_j}\) are not related.
The parameters \(a\) and \(b\) satisfy \(1 \ge a \ge b > 0\). For a particular dataset, the more trustworthy the tags are, the greater difference between \(a\) and \(b\) we set. In our experiments, we set \(a=1\) and \(b=0.1\).

\subsection{Learning to Hash with Multi-Level Topics}
Below, from different perspectives, we propose two strategies to integrate multi-granularity topics for improving short text hashing.

\begin{algorithm}[tb]\caption{Feature-Level Fusion Procedure}\label{alg:FeatureLevelFusion}
\begin{algorithmic}[1]
\renewcommand{\algorithmicrequire}{\textbf{Input:}}
\renewcommand{\algorithmicensure}{\textbf{Output:}}
\REQUIRE \
A set of \(n\) training texts \({\bf{X}} \) with tags \({\bf{t}} \), \(M\) optimal topic models \({\bf{O}} \) associated with their weight vector \({\vec{\hat \mu }} \).
\ENSURE \
The optimal hash codes \({\bf{Y}}\) and the hash function: \(l\) linear SVM classifiers.
\STATE Extract \(M\) topic feature sets \(\{ {{\vec{\theta }}_1},{{\vec{\theta }}_2},...,{{\vec{\theta }}_M}\} \) from the optimal topic models \({\bf{O}}\);
\STATE Produce the new feature \({\bf{\Omega }}\) by Eq.~\ref{eq:newFeature} and construct confidence matrix \({\bf{S }}\) by Eq.~\ref{eq:localSimialrity};
\STATE Obtain the \(l\)-dimensional vectors \({\bf{\tilde Y}}\) by optimizing Eq.~\ref{eq:TSH_Introduce};
\STATE Generate \({\bf{Y}}\) by thresholding \({\bf{\tilde Y}}\) to the median vector \({\bf{m}} = median({\bf{\tilde Y}})\);
\STATE Train \(l\) linear SVM classifiers by the learned codes \({\bf{Y}}\);
\RETURN Hash codes \({\bf{Y}}\) and \(l\) linear SVM;
\end{algorithmic}
\end{algorithm}
\vspace{-0.3cm}

\subsubsection{Feature-Level Fusion}
In order to integrate multi-granularity topics, we here adopt a simple but powerful way to combine observed features and latent features for short text, similar as \cite{ShortTextClassify_Phan_2008} and \cite{ShorTextMultiTopic_Chen_2011}, and create a high dimensional vector \({\bf{\Omega }}\) as:
\begin{equation}
\label{eq:newFeature}
{\bf{\Omega }} = [{\hat \mu _1}{{\vec{\theta }}_1},{\hat \mu _2}{{\vec{\theta }}_2},...,{\hat \mu _M}{{\vec{\theta }}_M}],
\end{equation}
where, \(\{ {{\bf{\theta }}_1},{{\bf{\theta }}_2},...,{{\bf{\theta }}_M}\} \) are the optimal topic features, and
\begin{equation}
\label{eq:unitWeight}
{\hat \mu _i} = {\mu _i}({T_i})/mi{n_{{T_k} \in { {\vec O}}}}({\mu _k}({T_k})).
\end{equation}

We can straightforwardly construct the similarity matrix \({\bf{S}}\) by Eq.~\ref{eq:localSimialrity} with the new features \({\bf{\Omega }}\) of training texts. Similar as Two-Step Hashing (TSH)~\cite{TwoStepHash_Lin_2013}, we see the binary code generation and hash function learning process as two separate steps. As a special example, Laplacian affinity loss and linear SVM are chosen to solve our problem. In first step, the training hash codes procedure can be formulated as following optimization:
\begin{equation}
\label{eq:TSH_Introduce}
\begin{array}{l}
\mathop {\min }\limits_{\bf{Y}} {\rm{ }}\sum\limits_{i,j = 1}^n {{S_{ij}}\left\| {{{\bf{y}}_i} - {{\bf{y}}_j}} \right\|_F^2} \\
s.t.{\rm{ }}{\bf{Y}} \in {\{  - 1,1\} ^{n \times l}},{{\bf{Y}}^T}{\bf{1}} = 0,{{\bf{Y}}^T}{\bf{Y}} = {\bf{I}}
\end{array}
\end{equation}
where \({S_{ij}}\) is the pairwise similarity between documents \({{\bf{x}}_i}\) and \({{\bf{x}}_j}\), \({{\bf{y}}_i}\) is the hash code for \({{\bf{x}}_i}\), and \({\left\|  \cdot  \right\|_F}\) is the Frobenius norm. To satisfy the similarity preservation, we seeks to minimize the quantity, because it incurs a heavy penalty if two similar documents are mapped far away. The problem is relaxed by discarding \({\bf{Y}}   \in   {\{  - 1,1\} ^{n \times l}}\), the optimal \(l\)-dimensional real-valued vector \({\bf{\tilde Y}}\) can be obtained by solving Laplacian Eigenmaps problem~\cite{LaplacianEigen_Belkin_2003}. Then, \({\bf{\tilde Y}}\) can be converted into binary codes \({\bf{Y}}\) via the media vector \({\bf{m}}   =   median({\bf{\tilde Y}})\). In hash function learning step, thinking of each bit \(y_i^{(p)}   \in   \{  + 1, - 1\} \) in the binary code as a binary class label for that text, we can train \(l\) linear SVM classifiers \(f({\bf{x}})   =   sgn({{\bf{W}}^T}{\bf{x}})\) to predict the \(l\)-bit binary code for any query document \({{\bf{x}}_q}\). Algorithm~\ref{alg:FeatureLevelFusion} shows the procedure of this strategy.

\begin{algorithm}[t]\caption{Decision-Level Fusion Procedure}\label{alg:DecisionLevelFusion}
\begin{algorithmic}[1]
\renewcommand{\algorithmicrequire}{\textbf{Input:}}
\renewcommand{\algorithmicensure}{\textbf{Output:}}
\REQUIRE \
A set of \(n\) training texts \({\bf{X}} \) with tags \({\bf{t}} \), \(M\) optimal topic models \({\bf{O}} \) and trade-off parameters, \({C_1}\) and \({C_2}\).
\ENSURE \
The optimal hash codes \({\bf{Y}}\) and a set of linear hash function matrices \({\bf{\tilde W}} \).
\STATE Extract \(M\) topic feature sets \(\{ {{\vec{\theta }}_1},{{\vec{\theta }}_2},...,{{\vec{\theta }}_M}\} \) from the optimal topic models \({\bf{O}}\);
\STATE Construct a series of confidence matrices \(\{ {{\bf{S}}^{(1)}},{{\bf{S}}^{(2)}},...,{{\bf{S}}^{(M )}}\} \) by Eq.~\ref{eq:localSimialrity} for \(M\) feature sets: \(\{ {{\vec{\theta }}_1},{{\vec{\theta }}_2},...,{{\vec{\theta }}_M}\} \);
\STATE Obtain the \(l\)-dimensional vectors \({\bf{\tilde Y}}\) and \({\bf{\tilde W}}\) by optimizing Eq.~\ref{eq:MFH_Objective};
\STATE Generate \({\bf{Y}}\) by thresholding \({\bf{\tilde Y}}\) to the median vector \({\bf{m}} = median({\bf{\tilde Y}})\);
\RETURN Hash codes \({\bf{Y}}\) and hash function matrix set \({\bf{\tilde W}}\);
\end{algorithmic}
\end{algorithm}
\vspace{-0.3cm}

\subsubsection{Decision-Level Fusion}
From another perspective, we can treat the optimal multi-granularity topic feature sets \(\{ {{\vec{\theta }}_1},{{\vec{\theta }}_2},...,{{\vec{\theta }}_M}\} \) extracted from short texts as multi-view features. In our situation, there are \(M\)-view features: \(\{ {{\vec{\theta }}_1},{{\vec{\theta }}_2},...,{{\vec{\theta }}_M}\} \). We take a linear sum of those \(M\)-view similarities as follows:
\begin{equation}
\label{eq:DecisionLevelFusion}
\sum\limits_{k = 1}^{M} {\sum\limits_{i,j = 1}^n {S_{ij}^{(k)}\left\| {{{\bf{y}}_i} - {{\bf{y}}_j}} \right\|_F^2} }
\end{equation}
where, \(S_{ij}^{(k)}\) constructed as Eq.~\ref{eq:localSimialrity} is the affinity matrix defined on the \(k\)-th view features. By introducing a diagonal \(n \times   n\) matrix \({{\bf{D}}^{(k)}}\) whose entries are given by \(D_{ii}^{^{(k)}}   =   \sum\nolimits_{j = 1}^n {S_{ij}^{(k)}} \), Eq.~\ref{eq:DecisionLevelFusion} can be rewritten as \(tr({{\bf{Y}}^T} \sum\limits_{k = 1}^{M}  {({{\bf{D}}^{(k)}}  -  {{\bf{S}}^{(k)}})} {\bf{Y}})  = tr({{\bf{Y}}^T} \sum\limits_{k = 1}^{M}  {{{\bf{L}} ^{(k)}}} {\bf{Y}})\), where \({{\bf{L}}^{(k)}}\) is the Laplacian matrix defined on the \(k\)-th view features. By introducing Composite Hashing with Multiple Information Sources (CHMIS)~\cite{CHMIS_Zhang_2011}, as a representative of Multiple View Hashing (MVH), we can simultaneously learn the hash codes \(\bf Y\) of the training texts \(\bf X\) as well as a set of linear hash functions \(\sum\nolimits_{k = 1}^M {{\alpha _k}{{({{\bf{W}}^{(k)}})}^T}{{\bf{X}}^{(k)}}} \) to infer the hash code for query text \({{\bf{x}}_q}\). The overall objective function is given as follows:
\begin{equation}
\label{eq:MFH_Objective}
\begin{array}{l}
\mathop {\min }\limits_{{\bf{Y}},{\bf{W}},{\bf{\alpha }}} {\rm{ }}{C_1}tr({{\bf{Y}}^T}\sum\limits_{k = 1}^{M} {{{{\bf{\tilde L}}}^{(k)}}} {\bf{Y}})
 + {C_2}\left\| {{\bf{Y}}  -   \sum\limits_{k = 1}^{M }  {{\alpha _k}} ({{\bf{W}}^{(k)}}){{\bf{X}}^{(k)}}} \right\|_F^2   +   \sum\limits_{k = 1}^{M}  {\left\| {{{\bf{W}}^{(k)}}} \right\|_F^2} \\
s.t.{\rm{  }}{\bf{Y}}   \in   {{\rm{\{  - 1,1\} }} ^{n \times k}}  ,{\rm{ }}{{\bf{Y}}^T}{\bf{1}}   =   0,{\rm{ }}{{\bf{Y}}^T}{\bf{Y}}   =   {\bf{I}}{\rm{, }}{{\bf{\alpha }} ^T}{\bf{1}}{\rm{  =   1, }}{\bf{\alpha }}   \ge   {\rm{0}}
\end{array}
\end{equation}
where, \({C_1}\) and \({C_2}\) are trade-off parameters, \(tr( \cdot )\) is the matrix trace function, \({\vec{\alpha }} = [{\alpha _1},{\alpha _2},...,{\alpha _M}]\) is a combination coefficient vector to balance the outputs from each view features, and a series of linear hash function matrices: \({\bf{\tilde W}} = \{ {\alpha _1}{{\bf{W}}^{(1)}},{\alpha _2}{{\bf{W}}^{(2)}},...,{\alpha _{M }}{{\bf{W}}^{(M )}}\}\). In order to solve this hard optimization problem, we first relax the discrete constraints \({\bf{Y}}   \in   {\{  - 1,1\} ^{n \times l}}\), and iteratively optimize one variable with the other two fixed. More detailed optimization procedures of this method can be found in~\cite{CHMIS_Zhang_2011}. Different from the former strategy, we do not need to pre-allocate the weight value of each view features, because that the combination coefficient vector \({\vec{\alpha }} = [{\alpha _1},{\alpha _2},...,{\alpha _M}]\) learned iteratively in the process of optimization can balance the outputs of each view features, and the procedure of this strategy is shown in Algorithm~\ref{alg:DecisionLevelFusion}.

\subsection{Complexity Analysis}
The training processes including binary code learning and hash function training are always conducted off-line. Thus, our focus of efficiency is on the prediction process. This process of generating hash code for a query text only involves some Gibbs sampling iterations to extract multi-granularity topics \(\{ {{\vec{\theta }}_1},{{\vec{\theta }}_2},...,{{\vec{\theta }}_M}\} \) and dot products in hash function \({\bf{y}} = sgn({{\bf{W}}^T}{\bf{x}})\), which can be done in \(O{\rm{(}}r\tilde Ks + l\tilde K{\rm{)}}\). Here, \(r\) is the number of Gibbs sampling iterations for topic inference, \(\tilde K\) is the sum of multi-granularity topic numbers \(\{ {K_1},{K_2},...,{K_M}\} \), \(l\) is the dimensionality of hash code and \(s\) denotes the sparsity of the observed keyword features. The values of the parameters above can be regarded as quite small constants. For example, \(r = 20,{\rm{ }}\tilde K \approx 100,{\rm{ }}l \le 64\) and the average number of sparsity per document \(s\) is no more than 100 in our experimental datasets. We can see the major time complexity is the Gibbs sampling for topic inference. In recent works, lots of studies focus to accelerate the topic inference. For example, in Biterm Topic Model (BTM), \cite{cheng2014btm} gives a simplicity and efficient method without Gibbs sampling iterations and the time complexity for topic inference can be reduced to \(O{\rm{(}}Kb{\rm{)}}\), where \(b\) is the number of biterms in a query text.

\section{Experiment and Analysis}
\label{sec:ExperimentAndAnalysis}
\subsection{Dataset and Experimental Settings}
We carried out extensive experiments on two publicly available real-world text datasets: one is typical short text dataset, {\em Search Snippets}\footnote{http://jwebpro.sourceforge.net/data-web-snippets.tar.gz}, and another is normal text dataset, {\em 20Newsgroups}\footnote{http://people.csail.mit.edu/jrennie/20Newsgroups/}.

The {\bf Search Snippets} dataset collected by Phan~\cite{ShortTextClassify_Phan_2008} was selected from the results of web search transaction using predefined phrases of 8 different domains. We further filter the stop words and stem the texts. 20139 distinct words, 10059 training texts and 2279 test texts are left, and the average text length is 17.1.

The {\bf 20Newsgroups} corpus was collected by Lang~\cite{20Newsgroups_Lang_1995}. We use the popular `bydate` version which contains 20 categories, 26214 distinct words, 11314 training texts and 7532 test texts, and the average text length is 136.7.

For these datasets, we denote the category labels as tags. For {\em Search Snippets}, we use a large-scale corpus~\cite{ShortTextClassify_Phan_2008} crawled from Wikipedia to estimate the topic models, and the original keyword features are directly used for learning the candidate topic models for {\em 20Newsgroups} due to the sufficient keyword features.
In order to evaluate our method's performance, we compute standard retrieval performance measures: recall and precision, by using each document in the test set as a query to retrieve documents in the training set within a specified Hamming distance. For the original keyword feature space cannot well reflect the semantic similarity of documents, even worse for short text, we simply test if the two documents share any common tag to decide whether a semantic similar text. This methodology is used in SH~\cite{SemanticHashing_Salakhutdinov_2009}, STH~\cite{STH_Zhang_2010}, CHMIS~\cite{CHMIS_Zhang_2011} and SHTTM~\cite{SHTTM_Wang_2013}.

Five alternative hashing methods compared with our proposed approach are STHs~\cite{STHs_Zhang_2010}, STH~\cite{STH_Zhang_2010}, LCH~\cite{LCH_Zhang_2010}, LSI~\cite{SemanticHashing_Salakhutdinov_2009} and SpH~\cite{SpH_Weiss_2008}. The results of all baseline methods are obtained by the open-source implementation provided on their corresponding author's homepage. In order to distinguish the proposed two strategies in our approach, the feature level fusion method is denoted as HMTT-Fea, and the decision level fusion method is named as HMTT-Dec\footnote{https://github.com/jacoxu/short-text-hashing-HMTT, http://www.CICLing.org/2015/data/148}.

In our experiments, the candidate topic sets {\bf{T}} = \{\( T10\), \( T30\), \( T50\), \( T70\), \( T90\), \( T120\), \( T150\)\} and the number of the optimal topic sets is fixed to 3. The parameters \({C_1}\) and \({C_2}\) in Eq.~\ref{eq:MFH_Objective} are tuned from \{0.1, 1, 10, 100\}. The number of nearest neighbors is fixed to 25 when constructing the graph Laplacians in our approach, as well as in the baseline methods, STHs and STH. We evaluate the performance of different methods by varying the number of hashing bits from 4 to 64. For LDA, we used the open-source implementation GibbsLDA\footnote{http://jgibblda.sourceforge.net/}, and the hyper-parameters are tuned as \(\alpha  = 0.5\), \(\beta  = 0.01\), 1000 iterations of Gibbs sampling for learning, and 20 iterations for topic inference. The results reported are the average over 5 runs.

\subsection{Results and Analysis}
We sample 100 texts for each category with tags information randomly from training dataset and set \(k\) in Eq.~\ref{eq:updateWeightU} to 10 to evaluate the quality of topic sets by Algorithm~\ref{alg:weightTopic}.
As the number of optimal topic sets is fixed to 3, we get the optimal topic sets {\bf{{\rm O}}} = \{\( T10\), \( T30\), \( T50\)\} for both two datasets coincidentally, and the weight vectors \({\bf{\hat \mu }}\) = \{3.44, 1.7, 1\} for {\em Search Snippets} and \({\bf{\hat \mu }}\) = \{1.31, 1.22, 1\} for {\em 20Newsgroups}. It is noteworthy that the weight values of the topic sets are affected by both the type of dataset and the settings of LDA.
Below, a series of experiments are conducted to answer the questions: (1). How does the proposed approach HMTT compare with other baseline methods; (2). Whether the optimal multi-granularity topics can outperform single-granularity topics and other multi-granularity topics; (3). Which approach of the two strategies to integrate multi-granularity topics can achieve a better performance.

\begin{figure}[tb]
\centering
\subfigure[SearchSnippets]{
\includegraphics[width=5.2cm]{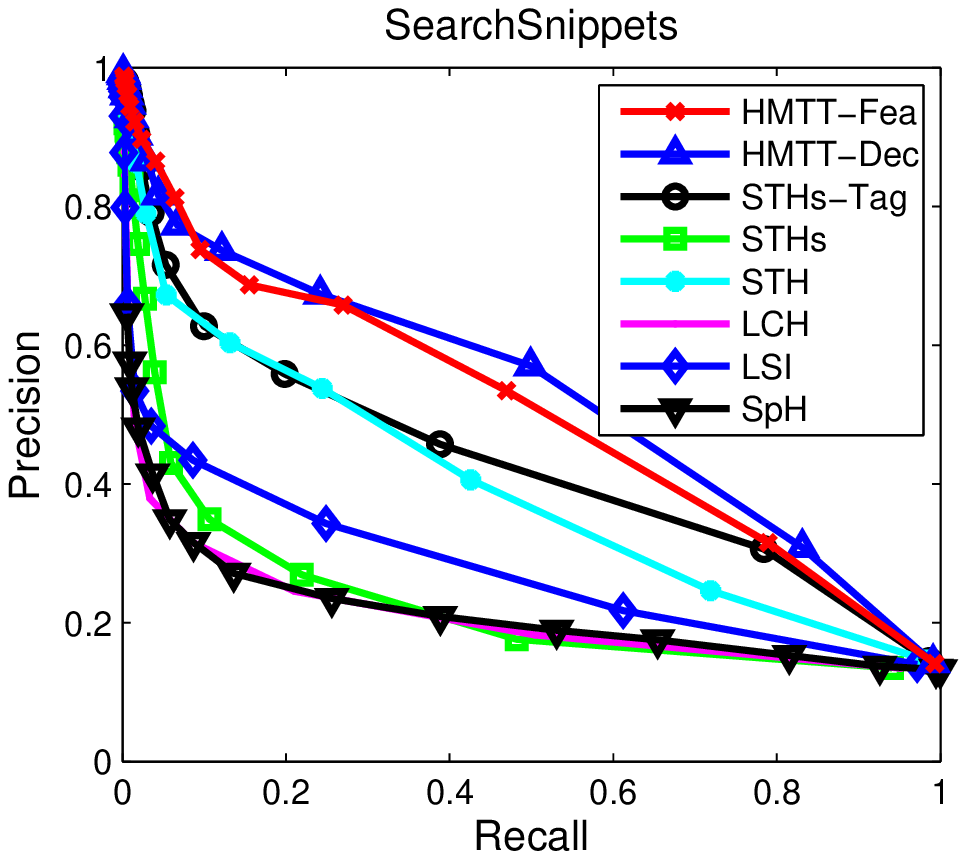}}
\subfigure[20Newsgroups]{
\includegraphics[width=5.2cm]{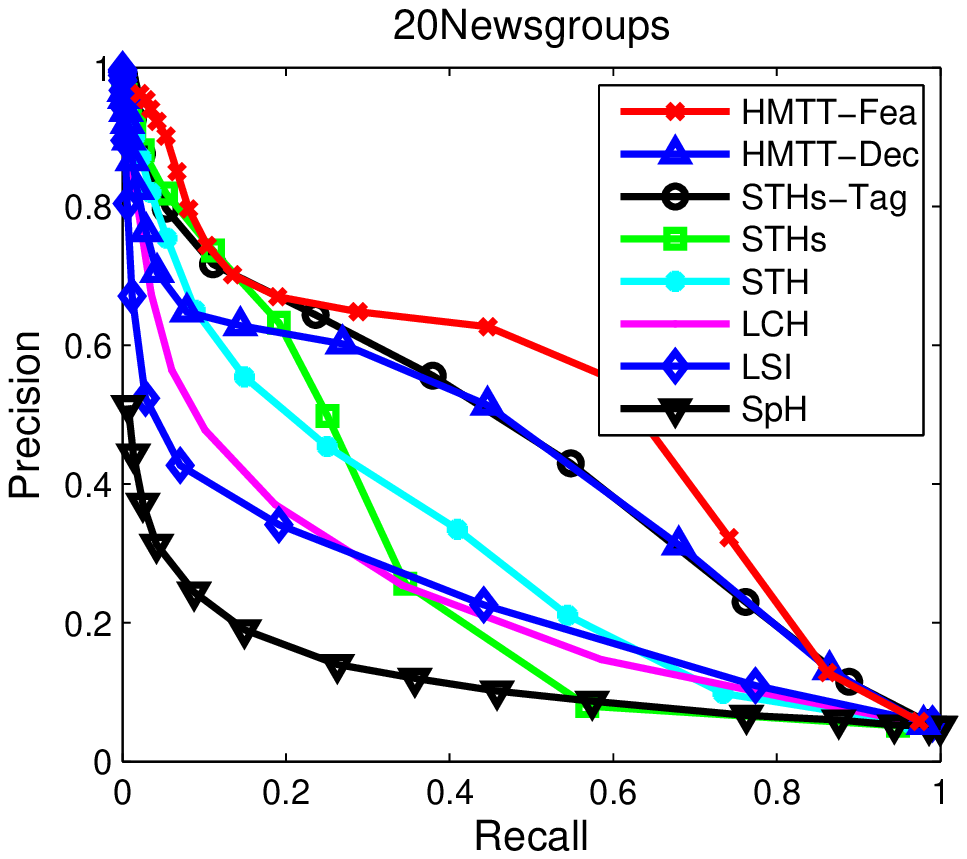}}
\vspace{-0.3cm}
\caption{Precision-Recall curves of retrieved examples within Hamming radius 3 on two datasets with different hashing bits (4:4:64 bits).}
\label{fig:PRonTwoDatasetWithDifferentBits}
\vspace{-0.5cm}
\end{figure}

{\bf{Compared with the existing hashing methods:}} In this section, we design an improved version of STHs, denoted as STHs-Tag, by replacing the original construction of similarity matrix with the proposed method described in Section~\ref{subsec:TagPreserve}. We remove 60 percent tags randomly from the training dataset to verify the robustness for HMTT-Fea, HMTT-Dec, STHs and STHs-Tag. The precision-recall curves for retrieved examples are reported in Fig.~\ref{fig:PRonTwoDatasetWithDifferentBits}.
From these comparison results, we can see that HMTT-Fea and HMTT-Dec significantly outperform other baseline methods on {\em Search Snippets} as shown in Fig.~\ref{fig:PRonTwoDatasetWithDifferentBits} (a). For {\em 20Newsgroups}, HMTT-Dec performs close results with STHs-Tag in Fig.~\ref{fig:PRonTwoDatasetWithDifferentBits} (b). The reasons to explain this problem are that: Firstly, {\em 20Newsgroups} as a normal dataset has sufficient original features to learn hash codes so that STHs-Tag based on keyword features works well. Secondly, we directly learn the topic models of {\em 20Newsgroups} from the training dataset that result in some restrictions. Furthermore, STHs get a worse performance than STHs-Tag on two datasets. Because STHs uses a complete supervised approach which only utilizes the pairwise similarity of the documents with common tags, that method cannot well deal with the situations that tags are missing or incomplete. In our approach, we extract the optimal multi-granularity topics depending on the type of dataset to learn hash codes and hashing function, and the tags are just utilized to adjust the similarity, which has stronger robustness. In the following experiment sets, we keep the all tags to improve the performance of hashing learning.

\begin{table*}[tb] 
\caption{\label{tb:PRonTwoDatasetWith816and32bitsWithTop200OS} Mean precision (mP) of the top 200 examples and the retrieved examples within Hamming radius 3 on {\em SearchSnippets} with 8 and 16 hashing bits. e.g. 10-30-50* means that the proposed methods incorporate the optimal multi-granularity topics, and 10-30-50W1 means that hashing method uses the multi-granularity topic sets \{\( T10\), \( T30\), \( T50\)\} while fixing the balance values to 1:1:1.}
\begin{center}
\begin{tabular}{ccccccccccccc}\hlinewd{1.2pt}
--- &  \multicolumn{4}{c}{mP@Top 200}&  \multicolumn{4}{c}{mP@Hamming Radius 3}\\ \hline
Methods & \multicolumn{2}{c}{HMTT-Fea} &  \multicolumn{2}{c}{HMTT-Dec} & \multicolumn{2}{c}{HMTT-Fea} &  \multicolumn{2}{c}{HMTT-Dec}\\\hline
Code Length&~8 bits~ &~16 bits~ &~8 bits~ &~16 bits~ &~8 bits~ &~16 bits~ &~8 bits~ &~16 bits~\\\hlinewd{1.2pt}
10-30-50* &\bf 0.829 &	0.799       &\bf 0.826 &\bf 0.782       &\bf 0.411   &\bf 0.802 &\bf	0.403 &\bf	0.778\\
10-70-90  &	0.819    &\bf 0.800     &	0.797  &	0.762       &	0.375    &	0.789  &	0.328     &	0.754\\
30-90-150 &	0.802    &	0.787       &	0.801  &	0.755       &	0.393    &	0.777      &	0.382     &	0.757\\
10-30     &	0.810    &	0.789       &	0.776  &	0.757       &	0.382    &	0.776      &	0.374     &	0.744\\
10-50     &	0.813    &	0.788       &	0.772  &	0.752       &	0.383    &	0.790      &	0.334     &	0.740\\
30-50     &	0.806    &	0.796       &	0.805  &	0.777       &	0.393    &	0.779      &	0.369     &	0.764\\
10-30-50W1&	0.811    &	0.780       &	0.822  &	0.778       &	0.368    &	0.761      &	0.398     &	0.774\\\hline
10        &	0.627    &	0.624       &	0.639  &	0.602       &	0.316&	0.610&	0.296&	0.576\\
30        &	0.792    &	0.764       &	0.728  &	0.708       &	0.377&	0.757&	0.335&	0.692\\
50        &	0.782    &	0.758       &	0.731  &	0.723       &	0.360&	0.730&	0.320&	0.707\\
70        &	0.771    &	0.755       &	0.728  &	0.720       &	0.365&	0.747&	0.318&	0.704\\
90        &	0.757    &	0.733       &	0.735  &	0.708       &	0.363&	0.736&	0.332&	0.692\\
120       &	0.730    &	0.705       &	0.707  &	0.700       &	0.366&	0.714&	0.309&	0.683\\
150       &	0.740    &	0.727       &	0.675  &	0.674       &	0.370&	0.729&	0.304&	0.660\\\hlinewd{1.2pt}
\end{tabular}
\end{center}
\vspace{-0.5cm}
\end{table*}

{\bf{Compared with single-granularity and other multi-granularity topic sets:}} Here, the hashing performances of the optimal multi-granularity topics are compared with single-granularity and other multi-granularity topics. We further evaluate the balance values of the multi-granularity topics by fixing them to 1. In particular, we keep the parameters \({\hat \mu _i}\) in Eq.~\ref{eq:newFeature} and \({\alpha _i}\) in Eq.~\ref{eq:MFH_Objective} to 1 for HMTT-Fea and HMTT-Dec respectively. The quantitative results on {\em Search Snippets} are reported in Table~\ref{tb:PRonTwoDatasetWith816and32bitsWithTop200OS}. From the results, we can see that the performances of multi-granularity topics significantly outperform single-granularity topics and the optimal multi-granularity topics achieve a better performance in most situations. We also observe similar results on {\em 20Newsgroups}. But due to the limit of space, we select to present the results on the typical short texts dataset {\em Search Snippets}.

{\bf{Compared between the proposed two strategies:}} Finally, we mainly discuss the performances between the proposed two strategies, HMTT-Fea and HMTT-Dec. In HMTT-Fea, we directly concatenate the multi-granularity topics to produce one feature vector and decompose the hashing learning problem into two separate stages. In HMTT-Dec, the multi-granularity topics extracted from the text content are treated as multi-view features, and we simultaneously learn the hash codes as well as hash function. From the results in Table~\ref{tb:PRonTwoDatasetWith816and32bitsWithTop200OS}, we can see that the performances of HMTT-Fea surpass HMTT-Dec on several evaluation metrics. Obviously, the former strategy is more simple and effective for short text hashing in our approach. In summary, no matter in HMTT-Fea or HMTT-Dec, the experimental results indicate that short text hashing can be improved by integrating multi-granularity topics.

\section{Discussions and Conclusions}
\label{sec:DiscussionANDConclusions}

Short text hashing is a challenging problem due to the sparseness of text representation. In order to address this challenge, tags and latent topics should be fully and properly utilized to improve hashing learning. Furthermore, it is better to estimate the topic models from an external large-scale corpus and the optimal topics should be selected depending on the type of dataset. This paper uses a simple and effective selection methods based on symmetric KL-divergence of topic distributions, we think that there are many other selection methods worthy of being explored further. Another key issue worthy of research is how to integrate the multi-granularity topics effectively. In this paper, we propose a novel unified hashing approach for short text retrieval. In particular, the optimal multi-granularity topics are chosen depending on the type of dataset. We then use the optimal multi-granularity topics to learn hash codes and hashing function on two distinct ways, meanwhile, tags are utilized to enhance the semantic similarity of related texts. Extensive experiments demonstrate that the proposed method can perform better than the competitive methods on two public datasets.

\section*{Acknowledgments}
This work is supported by the National Natural Science Foundation of China under Grant No. 61203281 and No. 61303172.

\bibliographystyle{splncs03}
\bibliography{sigproc}  

\end{document}